# Magnetic Lattice Dynamics of the Oxygen-Free FeAs Pnictides: How Sensitive are Phonons to Magnetic Ordering?

**Mohamed Zbiri**[1], **Ranjan Mittal**[2,3], **Stéphane Rols**[1], **Yixi Su**[2], **Yinguo Xiao**[4], **Helmut Schober**[1,5], **Samrath L. Chaplot**[3], **Mark R. Johnson**[1], **Tapan Chatterji**[6], **Yasunori Inoue**[7], **Satoru Matsuishi**[7], **Hideo Hosono**[7] and **Thomas Brueckel**[2,4]

[1] Laue Langevin Institute, 38042 Grenoble Cedex 9, France
[2] Juelich Centre for Neutron Science, IFF, Forschungszentrum Juelich, Outstation at FRM II, Lichtenbergstr. 1, D-85747 Garching, Germany
[3] Solid State Physics Division, Bhabha Atomic Research Centre, Trombay, Mumbai 400 085, India
[4] Institut fuer Festkoerperforschung, Forschungszentrum Juelich, D-52425 Juelich, Germany
[5] Université Joseph Fourier, UFR de Physique, 38041 Grenoble Cedex 9, France
[6] Juelich Centre for Neutron Science, Forschungszentrum Juelich, Outstation at Institut Laue-Langevin, BP 156, 38042 Grenoble Cedex 9, France
[7] Frontier Research Center, Tokyo Institute of Technology, 4259 Nagatsuta-cho, Midori-ku, Yokohama 226-8503, Japan

E-mail: `zbiri@ill.fr`

**Abstract.** To shed light on the role of magnetism on the superconducting mechanism of the oxygen-free FeAs pnictides, we investigate the effect of magnetic ordering on phonon dynamics in the low-temperature orthorhombic parent compounds, which present a spin-density wave. The study covers both the 122 ($AFe_2As_2$; A=Ca, Sr, Ba) and 1111 (AFeAsF; A=Ca, Sr) phases. We extend our recent work on the Ca (122 and 1111) and Ba (122) cases by treating computationally and experimentally the 122 and 1111 Sr compounds. The effect of magnetic ordering is investigated through detailed non-magnetic and magnetic lattice dynamical calculations. The comparison of the experimental and calculated phonon spectra shows that the magnetic interactions/ordering have to be included in order to reproduce well the measured density of states. This highlights a spin-correlated phonon behavior which is more pronounced than the apparently weak electron-phonon coupling estimated in these materials. Furthermore, there is no noticeable difference between phonon spectra of the 122 Ba and Sr, whereas there are substantial differences when comparing these to $CaFe_2As_2$ originating from different aspects of structure and bonding.

# 1. Introduction

The discovery of superconductivity in the FeAs compounds with high superconducting temperatures has led to considerable activity in the field of condensed matter physics. Extensive studies have been conducted to clarify the mechanism of Cooper pair formation and to improve Tc values. Since then, mainly three types of FeAs-based superconductors have been discovered: RFeAsO [1] (R: Rare earth elements) or AFeAsF (A: alkaline-earth elements), AFe$_2$As$_2$ [2, 3, 4], and LiFeAs [5]. For 122 compounds the highest Tc was obtained [2, 3] in K-doped BaFe$_2$As$_2$ (Tc=38 K). However, a much higher Tc of 56 K and 57.4 K has been found for Sr$_{0.5}$Sm$_{0.5}$FeAsF [6] and Ca$_{0.4}$Nd$_{0.6}$FeAsF [7], respectively. The 1111 oxygen based compounds SmFeAsO$_{0.9}$F$_{0.1}$ [8] and Gd$_{0.8}$Th$_{0.2}$FeAsO [9] also have high Tc of 55 K and 56 K respectively.

Structural, electronic and magnetic properties of these compounds are remarkably similar. Structural investigations indicate that these compounds have a tetragonal phase [2, 3, 10] at room temperature. The parent FeAs pnictides show a well pronounced spin density wave through transitions from the paramagnetic, tetragonal phase to an antiferromagnetically ordered, orthorhombic phase [2, 3, 10, 11]. Consequently, magnetism couples to the structure in these systems. In AFe$_2$As$_2$ (A=Ba, Ca and Sr) electron/hole doping or pressure suppresses these phase transitions and the related magnetic ordering and induces superconductivity through electron pairing at lower temperatures, supporting the idea of a possible coupling between spin degrees of freedom and superconductivity in the iron pnictides.

Inelastic neutron scattering measurements have evidenced a resonant spin excitation in polycrystalline Ba$_{0.6}$K$_{0.4}$Fe$_2$As$_2$ [12] as well as in single crystals of Ba(Fe$_{0.92}$Co$_{0.08}$)$_2$As$_2$ [13] and BaFe$_{1.9}$Ni$_{0.10}$As$_2$ [14]. Recent experiments have established that the role of phonons in the electron pairing mechanism cannot be entirely discarded. In particular, the observation of a large Fe isotope effects was shown in SmFeAsO$_{0.85}$F$_{0.15}$ and Ba$_{0.6}$K$_{0.4}$Fe$_2$As$_2$ systems [15]. Further inelastic X-ray scattering experiments on parent SmFeAsO and superconducting SmFeAsO$_{0.65}$F$_{0.35}$ single crystals indicate that specific phonon modes are strongly renormalized upon fluorine doping [16]. Inelastic X-ray scattering measurements [17] on polycrystalline NdFeAsO$_{1-y}$F$_y$ samples also revealed softening in the phonon density of states under F doping.

The role of spin-phonon coupling has also been investigated experimentally by Raman scattering. The measurement of the temperature dependence of zone centre phonons modes in Ba$_{1-x}$K$_x$Fe$_2$As$_2$ [18] and Ba(Fe$_{1-x}$Co$_x$)$_2$As$_2$ [19] show renormalization of the zone center Raman active phonons on cooling through the tetragonal-to-orthorhombic transition temperature where the magnetic ordering occurs.

On the computational side, magnetism has been found to be of considerable importance to treat reasonably the oxygen-based and oxygen-free iron-pnictides. In a very short time a big amount of work has been done to highlight the discrepancies between non-magnetic and magnetic calculations to describe the Fe-pnictides (see for instance

Refs [20, 21, 22, 23, 24, 25, 26] and references therein). For a closely related topic to the presently reported one, ab-initio calculations for $BaFe_2As_2$ show evidence for a selective coupling of electronic and spin degrees of freedom with phonons [27]. Furthermore, the measurement of the phonon density of states for the Ca 1111 $CaFe_{1-x}Co_xAsF$ (x = 0, 0.06, 0.12) indicates [28] that stronger spin-phonon interactions play an important role for the emergence of superconductivity in these compounds. It is hence of a primary interest to investigate the direct effect of the spin degrees of freedom on the phonon dynamics in the parent compounds in the low temperature phase where the spin density wave occurs. In this context we consider magnetic ordering, observed in the orthorhombic phase, and probe spin-phonon coupling. Here we report detailed ab initio simulations where non magnetic calculations (NM) are used as a reference and compared to the magnetic case (MAG) by explicitly including the additional spin degree of freedom and the related ordering. Both the 122 ($AFe_2As_2$, A=Ca, Sr, Ba) and 1111 (AFeAsF, A=Ca, Sr) phases are considered. The calculations are accompanied by new observations which, in addition to the Ca and Ba cases [27, 28, 29, 30, 31, 32], extend the measurements of the temperature dependence of phonon spectra to the parent compounds $SrFe_2As_2$ (122) and SrFeAsF (1111).

The aim of the present work is (i) to establish from ab initio calculations the effect of the magnetism on phonon dynamics, (ii) investigate how sensitive phonons are to details of magnetostructural correlation, and (iii) establish if there is a systematic trend in the phonon dynamics upon substitution of the divalent cation A in 122 (A=Ca, Sr, Ba) and in 1111 (A=Ca, Sr), and how this cation change affects structure and dynamics in these materials.

This paper is organized as follows: The computational procedure for the magnetic lattice dynamics calculations is detailed in Section 2. The experimental details of the measurements of the Sr parent compounds 122 and 1111 are provided in Section 3. Section 4 is dedicated to the presentation and discussion of the results. A summary and some concluding remarks are drawn in Section 5.

## 2. Computational Details

Ab-initio calculations were performed using the Vienna ab initio simulation package (VASP) [33, 34], as described previously [27, 28, 29]. The effect of magnetic ordering, occurring in the low-temperature orthorhombic phase, on the simulated phonon density of states (PDS) and phonon dispersion relations (PDR) is investigated in detail here. NM calculations consist of neglecting the spin polarization while calculating the interatomic force constants of the dynamical matrix. In previous work these calculations have been based on either the experimentally determined unit cell or a cell optimized by including spin degrees of freedom, as experimentally observed; ferromagnetic along the b-direction and antiferromagnetic along the a and c axes [27]. It was found that either approach was required to maintain a reasonable value of the c-axis, that is the inter FeAs plane spacing. Unit cell optimization without spin polarization leads to a collapse of the inter-plane distance. In the MAG calculations, the spin degrees of freedom are included when calculating the inter-atomic force constants.. All phonon calculations are performed in the orthorhombic space groups, Fmmm (69) [$D_{23}^{2h}$] and Cmma (67) [$D_{21}^{2h}$] for the 122 and 1111 families, respectively, even when the magnetic order reduces this symmetry. As stated above, all spin polarized, ab initio calculations are performed with the observed magnetic ordering. This apparent discrepancy in treating symmetry is discussed later. It is worthwhile to notice that although the crystallographic properties in the Fe-pnictides are better described considering the magnetic interactions, the calculated magnetic moments are overestimated comparing to the measurements from neutron scattering and $\mu$SR experiments. This topic has been already discussed and was extensively studied. Different explanations were proposed ranging from frustration and (a non-observed) lattice distortion to spin fluctuations and effect of lattice-dimensionality (Refs [2, 27, 35, 36, 37, 38, 39] and references therein provide some relevant details on this regards for both measured and calculated magnetic moments in the Fe-pnictides).

In the lattice dynamics calculations, in order to determine all inter-atomic force constants, the super cell approach has been adopted [40, 41]. For 122, the single cell was used to construct a (2*a, 2*b, c) super cell containing 16 formula-units (80 atoms), a and b being the shorter cell axes. Total energies and inter-atomic forces were calculated for the 18 structures resulting from individual displacements of the three symmetry inequivalent atoms along the three Cartesian directions ($\pm$x, $\pm$y and $\pm$z).

In the MAG calculations of 1111, the single cell was used to construct a (2*a, 2*b, 2*c) super cell containing 32 formula-units (128 atoms). Total energies and inter-atomic forces were calculated for the 24 structures resulting from individual displacements of the four symmetry inequivalent atoms. NM calculations are performed without doubling along the c-axis leading to a (2*a, 2*b, c) super cell containing 16 formula-units (64 atoms). In this case 24 structures result from individual displacements of the four symmetry inequivalent atoms along the three Cartesian directions ($\pm$x, $\pm$y and $\pm$z). PDS and PDR, for both families, were extracted in subsequent calculations using the Phonon software [41].

## 3. Experimental Details of the Phonon Measurements of $SrFe_2As_2$ and $SrFeAsF$

We have previously reported detailed measurements of the temperature dependence of phonon spectra of $BaFe_2As_2$, $CaFe_2As_2$ and $CaFeAsF$ [28, 29, 30, 31, 32]. To allow systematic comparison in a broader series of compounds we have performed new measurements for the Sr compounds, which are reported here.

The polycrystalline samples of $SrFe_2As_2$ and $SrFeAsF$ were prepared [42] by heating stoichiometric mixtures of the corresponding purified elements. Structural analysis from x-ray powder diffraction indicates that the $SrFe_2As_2$ and $SrFeAsF$ samples contain about 1% of FeAs as an impurity phase. The inelastic neutron scattering experiments were performed using the IN4C and IN6 time of flight spectrometers at the Institut Laue Langevin (ILL), France as in our previous works [27, 28, 29]. Measurements were made on about 10 grams of polycrystalline samples. We have used incident neutron wavelengths of 1.2 $\mathring{A}$ (56.8 meV) and 5.12 $\mathring{A}$ (3.12 meV) for measurements on IN4C and IN6, respectively. The high incident neutron wavelength of 5.12 $\mathring{A}$ on IN6 only allows measurements to be performed in neutron-energy gain, so data can not be measured down to very low temperatures. The low incident neutron wavelength of 1.2 $\mathring{A}$ at IN4C allows the measurements of phonon spectra in the energy loss mode and therefore down to low temperatures ($\sim$2 K). The incoherent approximation [43, 44] has been used for extracting neutron weighted phonon density of states from the measured scattering function S(Q,E).

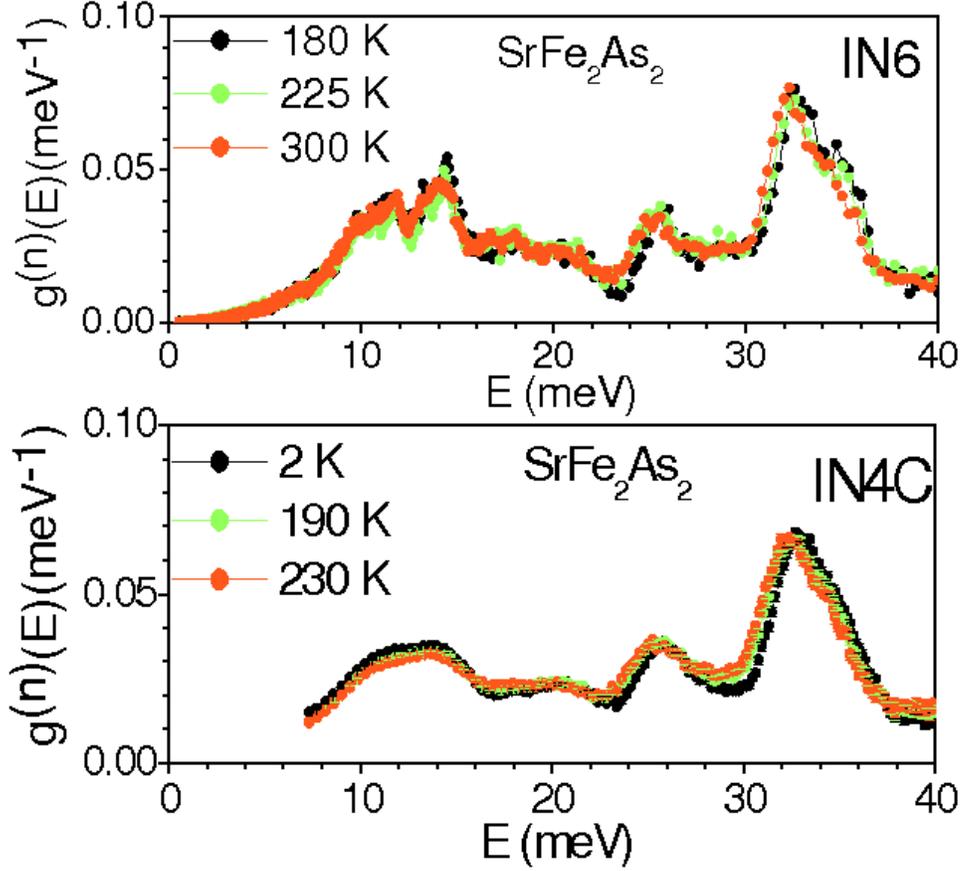

**Figure 1.** Temperature dependence of experimental phonon spectra for SrFe$_2$As$_2$. The phonon spectra are measured with incident neutron wavelength of 5.12 $\mathring{A}$ and 1.2 $\mathring{A}$ using the IN6 and IN4C spectrometers at the ILL, respectively.

## 4. Effect of Magnetic Ordering: Results and Discussion

We shall show first results of the new inelastic neutron scattering measurements of the parent Sr compounds (122 and 1111), which facilitate comparison with the Ca and Ba data previously reported [27, 28, 29]. Thereafter, the results of the magnetic lattice dynamics calculations are discussed in terms of magnetic ordering and structure-dependence upon change of the divalent cation A (Ca, Sr, Ba).

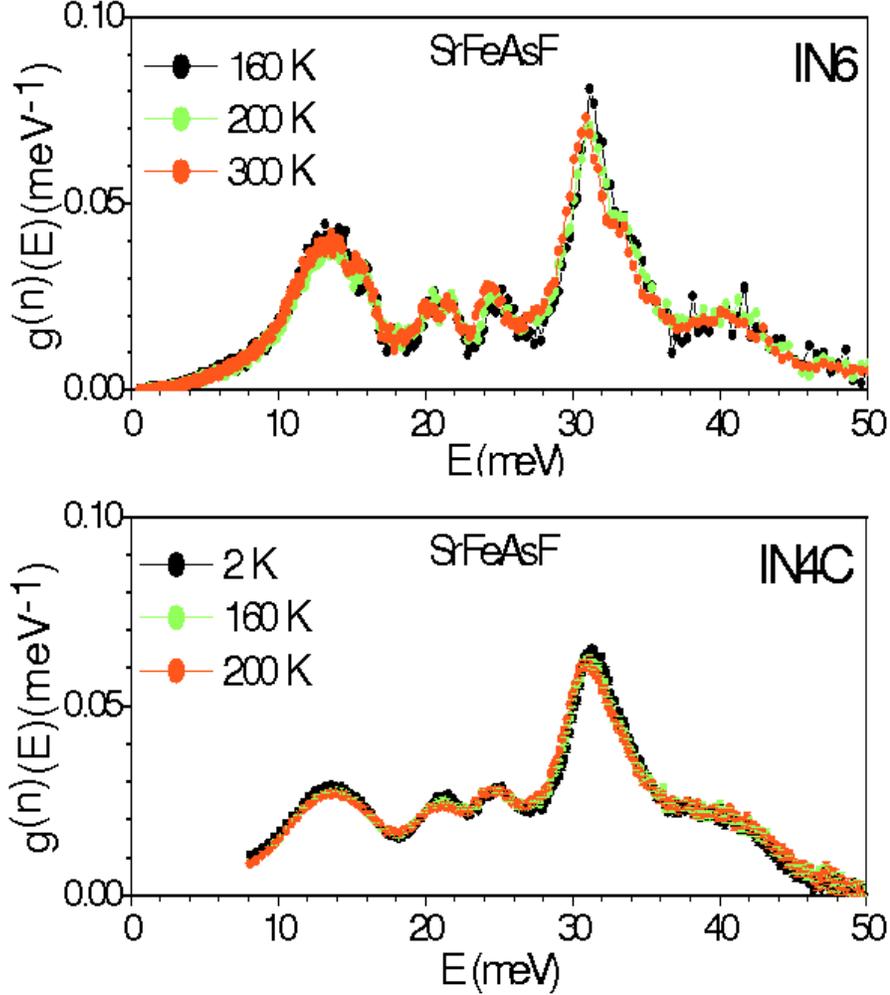

**Figure 2.** Temperature dependence of experimental phonon spectra for SrFeAsF. The phonon spectra are measured with incident neutron wavelength of 5.12 $\mathring{A}$ and 1.2 $\mathring{A}$ using the IN6 and IN4C spectrometers at the ILL, respectively.

### 4.1. SrFe₂As₂ and SrFeAsF: Experimental Results

The temperature dependence of the phonon spectra for the parent compound SrFe$_2$As$_2$ is reported in Figure 1. The orthorhombic to tetragonal phase transition occurs at 200

K [37, 45, 46]. The measurements at IN4C were carried out at 2 K, 190 K and 230 K, while the high resolution measurements at IN6 were performed at 180 K, 225 K and 300 K. The IN4C measurements show that low energy modes below 10 meV harden slightly as temperature is increased from 2 K to 190 K, while IN6 measurements show that modes below 22 meV do not vary with the temperature change from 150 K to 300 K. The peaks at 25 meV and 32 meV soften on heating in both IN4C and IN6 data and the 25 meV peak narrows.

SrFeAsF is the parent compound of the Sm-doped counterpart $Sr_{0.5}Sm_{0.5}FeAsF$ [6] having a very high Tc of 56 K. Our neutron diffraction measurements show that the parent compound SrFeAsF undergoes [47] a tetragonal to orthorhombic phase transition at 180 K followed by the magnetic phase transition at 133 K. Figure 2 shows the temperature dependence of the measured phonon spectra of SrFeAsF at 160 K, 200 K and 300 K using IN6, and at 2 K, 160 K and 200 K using IN4C. The phonon spectrum from 22 meV to 35 meV, including the peak at 32 meV due to stretching modes of Fe-As, shows softening with increase of temperature. The peak at 25 meV, due to Fe (see Figure 8 for calculated partial density of states) become narrower and moves slightly to higher energies on cooling from 300 K, contrary to the expectation of broadening due to the tetragonal to orthorhombic phase transition.

Our measurements show that the temperature variation across the tetragonal to orthorhombic phase transition or magnetic phase transition has little effect on the phonon dynamics of the Sr parent compounds.

## 4.2. Effect of magnetic ordering on phonon spectra

We first compare the observed phonon spectra as a function of the divalent cationic substitution (A=Ba, Sr, Ca). The high-resolution phonon density of states measured for the all $AFe_2As_2$ compounds using IN6 at 300 K are shown in Figure 3. There are pronounced differences in the low- and mid- frequency ranges below 22 meV, but there is no effect of changing the cation A (Ba, Sr, Ca) on the 25 meV peak. However as expected the Fe-As stretching modes around 32 meV shift towards lower energies in the order of Ba, Sr and Ca compounds due to the increase of the Fe-As bond lengths. All the 122 compounds have nearly the same value of the lattice parameter a ( 3.89 Å), while the lattice parameter c in Ca, Sr and Ba compounds is 11.758 Å, 12.37 Å and 13.04 Å, respectively. The lattice constant c decreases monotonically from $BaFe_2As_2$ to $SrFe_2As_2$ to $CaFe_2As_2$, as one would expect given the ionic radii: $Ca^{2+} < Sr^{2+} < Ba^{2+}$. Considering the large difference in the mass of Ba (m= 137.34 amu), Sr (m= 87.62 amu) and Ca (m=40.08 amu), the mass effect and the contraction of the unit cell should result in shifting of the phonon modes in the Ca compound to higher energies compared to the Ba and Sr cases. However, we find that the peaks at 12 meV and 21 meV in the Ba and Sr compounds are shifted to 10 meV and 19 meV in the Ca compound.

It is also worthwhile highlighting the importance of the peak at 21 meV in $BaFe_2As_2$. Experimentally it corresponds to an axially (z) polarized $A_{1g}$ mode of the As atoms. It has been shown that a small displacement of the As along the z-axis has a significant impact on the electronic and magnetic features of the BaFe₂As₂ [36, 38, 39]. This is clearly evident in the lattice dynamical calculations. The comparison between the experimental phonon spectra and ab-initio calculations for BaFe₂As₂ is shown in Figure. 4. We have calculated the density of states in the orthorhombic phase with (MAG) and without (NM) magnetic ordering. The NM calculations do not describe accurately the measured phonon spectrum. There is a pseudo gap at the 21 meV peak position and frequencies of the stretching modes are underestimated. Interestingly, MAG calculations recover all the observed features, especially the peak at 21 meV. An improvement is also observed for the Fe-As stretching modes. This is a clear signature of a magneto-structural correlation which affects phonon dynamics, and could indicate a selective spin-phonon coupling as recently suggested [27]. The situation is similar in the CaFe₂As₂ and SrFe₂As₂ cases shown in Figure 5, where magnetism improves the agreement.

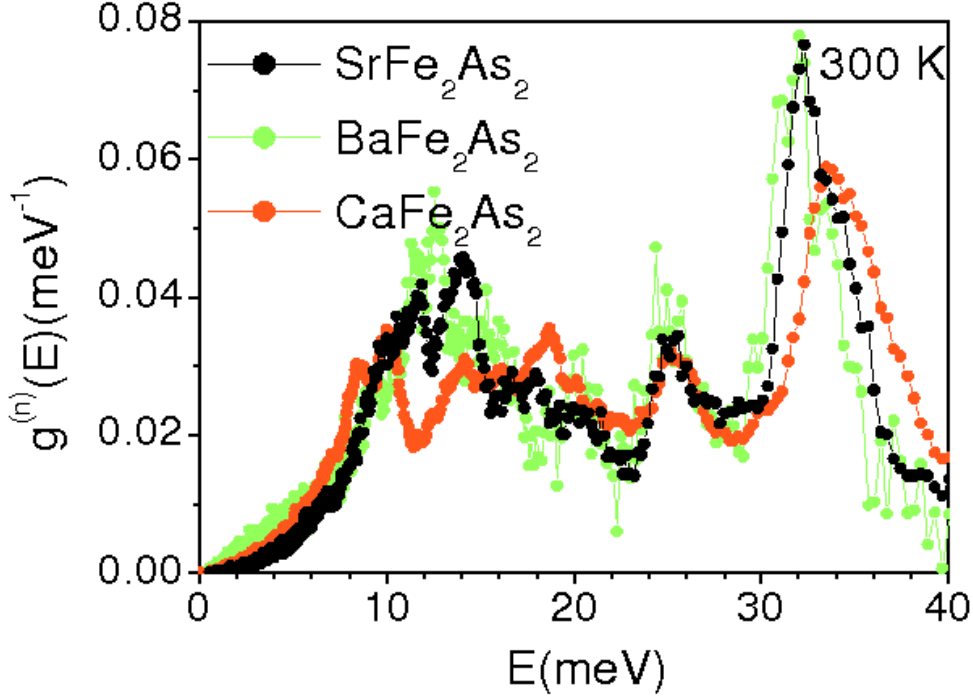

**Figure 3.** Comparison of the experimental phonon spectra for AFe₂As₂ (M=Sr, Ba, Ca). The phonon spectra are measured with an incident neutron wavelength of 5.12 Å using the IN6 spectrometer at the ILL. The experimental phonon data for BaFe₂As₂, CaFe₂As₂ are taken from Refs. [27] and [29], respectively. All the phonon spectra are normalized to unity.

Now we analyze the 1111 case by comparing the two systems CaFeAsF and SrFeAsF. Prior to discussing effect of magnetism, we compare first the high resolution phonon spectra of both compounds measured using the IN6 spectrometer. The two phonon spectra show pronounced differences in the full spectral range (Figure 6). At room temperature both compounds crystallize in a tetragonal structure (space group P4/nmm). The unit cell volume of CaFeAsF (a=3.8859 $\mathring{A}$, c=8.595 $\mathring{A}$) is about 10% smaller compared to SrFeAsF (a=3.9996 $\mathring{A}$ c=8.9618 $\mathring{A}$). It is not possible to attribute these changes to a simple change in unit cell volume and mass renormalization of the modes involving the Sr (m= 87.62 amu) and Ca (m=40.08 amu) atoms. Qualitatively, our data show in particular that the peaks at 10 meV and 16 meV in CaFeAsF are combined together in one peak centered at about 14 meV in SrFeAsF. Consequently a gap is opened up in the phonon density of states of SrFeAsF at about 18 meV. Another important change in the phonon spectra is the intensity of the peak at 32 meV. The peak has very high intensity in SrFeAsF compared to CaFeAsF. Further, the peak centered at about 45 meV in CaFeAsF is shifted to about 40 meV in SrFeAsF.

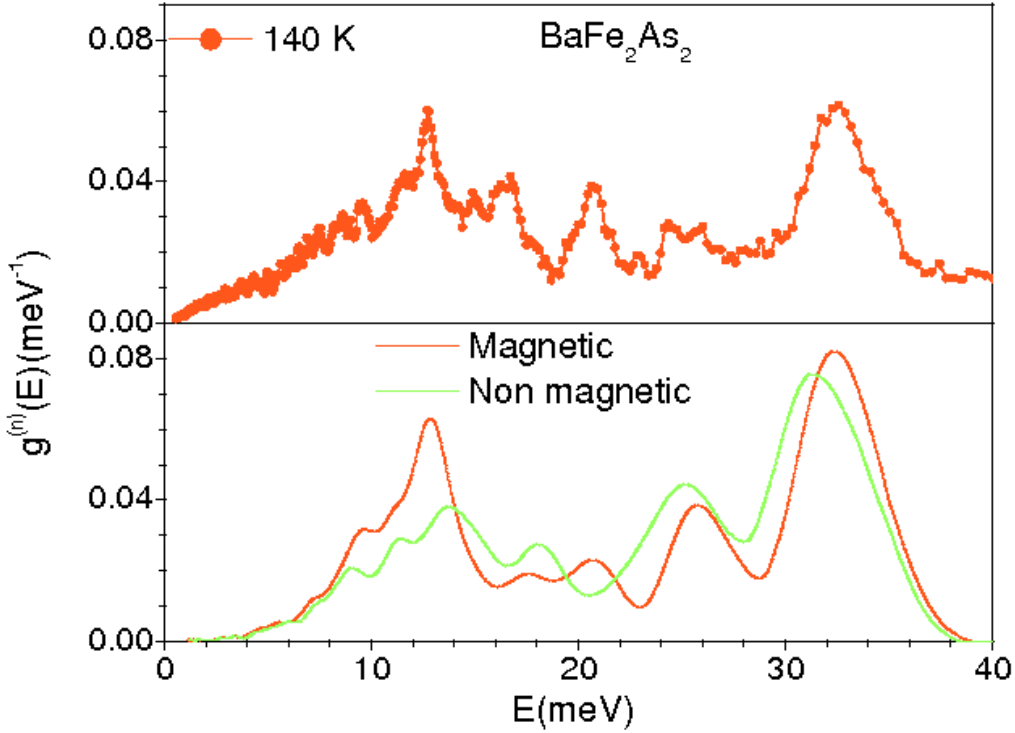

**Figure 4.** Comparison between the experimental and calculated phonon spectra for BaFe₂As₂. The calculated spectra have been convoluted with a Gaussian of FWHM of 10% of the energy transfer in order to describe the effect of energy resolution in the experiment. All the phonon spectra are normalized to unity. The experimental phonon data for BaFe₂As₂ are taken from Ref. [27].

Figure 7 compares the measured and calculated phonon spectra for the two 1111 systems. The observed features in the experimental data are well reproduced computationally when magnetism is considered, as it is the case for the 122 compounds. Both the peak positions and intensities are significantly improved, especially for the low-frequency and mid-frequency ranges. In SrFeAsF, the modes within the range 20-28 meV involving Fe and As are well resolved and match better the observations when magnetic interactions are included. Similarly, the low-lying 0 -12 meV frequency range is well described in CaFeAsF through a correct $E^2$-dependence in the MAG calculations. This is also the case for the intensity profile in 20  30 meV. Furthermore for both systems, the stretching modes due to F (shoulder around 45 meV) are found in the correct frequency position, indicating a good description of both structural and chemical interactions involving the lightest element F.

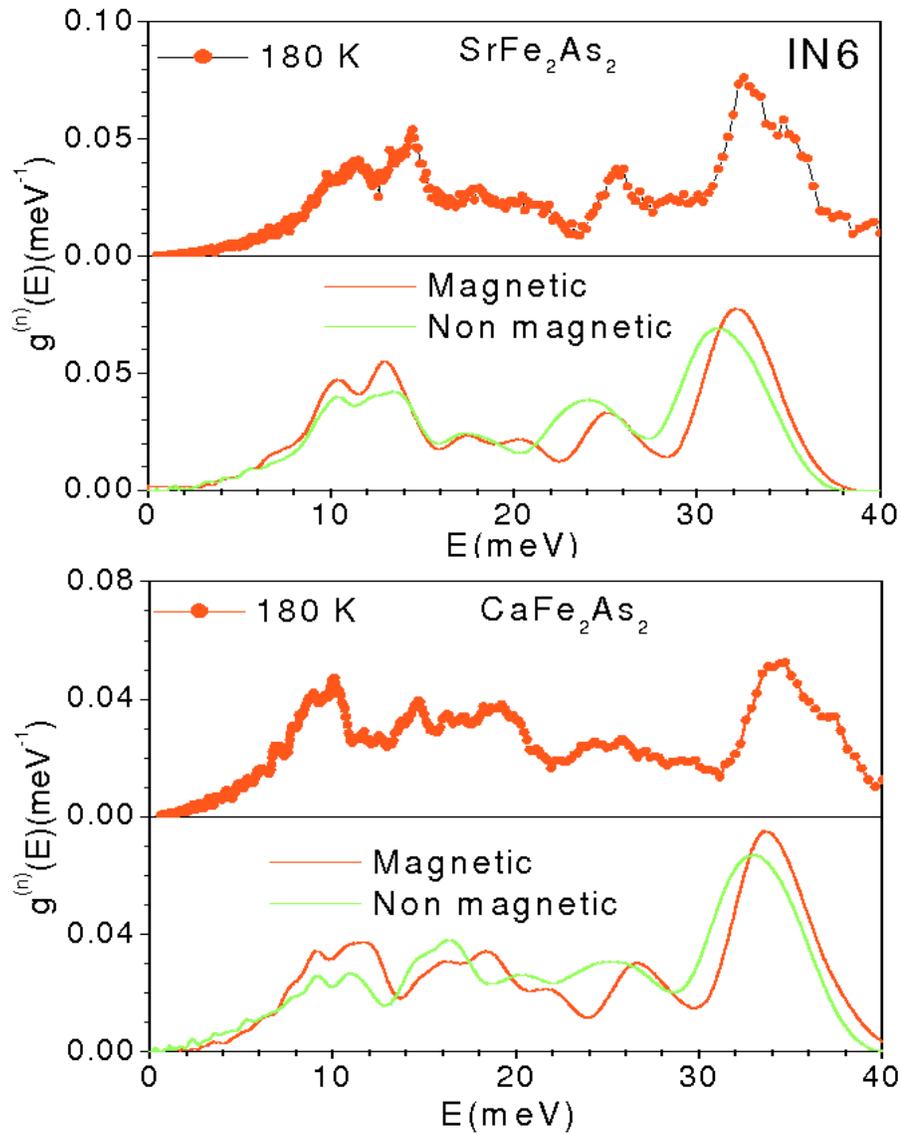

**Figure 5.** Comparison between the experimental and calculated phonon spectra for AFe$_2$As$_2$ (A=Sr, Ca). The calculated spectra have been convoluted with a Gaussian of FWHM of 10% of the energy transfer in order to describe the effect of energy resolution in the experiment. All the phonon spectra are normalized to unity. The experimental phonon data for CaFe$_2$As$_2$ are taken from Ref. [29]

For all the compounds (122 and 1111), the importance of the magnetic interactions to describe accurately the inter-atomic force constants is clearly demonstrated. The observed magnetic interactions should be included in the lattice dynamical calculations to describe accurately the phonons in the FeAs pnictides.

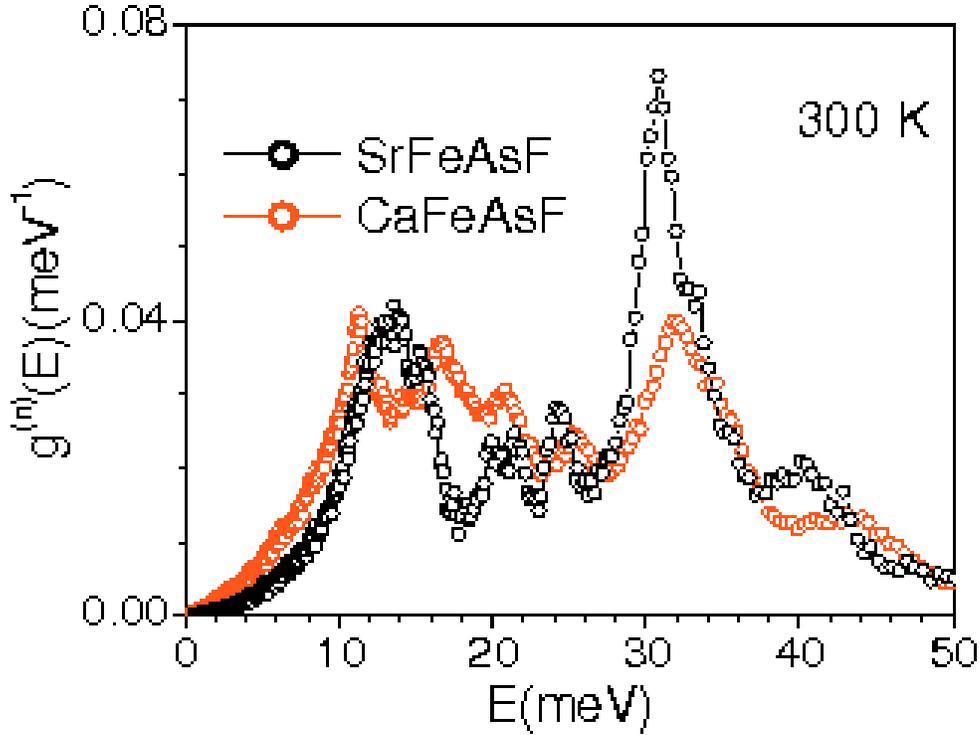

**Figure 6.** Comparison of the experimental phonon spectra for SrFeAsF and CaFeAsF. The phonon spectra are measured with an incident neutron wavelength of 5.12 Å using the IN6 spectrometer at the ILL. The experimental phonon data for CaFeAsF are taken from Ref. [28]. All the phonon spectra are normalized to unity.

*4.3. Further Considerations: Magnetically Calculated Partial Density of States*

Figure 8 shows the magnetically calculated (MAG) partial density of states for all the 122 (left panel) and 1111 (right panel) compounds presently discussed.

For the 122 cases, we find that the vibrational modes due to A (Ba, Sr, Ca) atoms in $AFe_2As_2$ scale approximately in the ratio expected from the masses of the Ba, Sr and Ca atoms. Thus their behavior is correctly reproduced. Furthermore, the Fe and As vibrations in Ba and Sr compounds are similar. However we observe that there is a substantial difference in the vibrations of Fe and As atoms in Ca compounds. The calculated Fe and As vibrations are found to soften in the range below 20 meV by about 1 to 2 meV in the Ca compound compared to the Ba and Sr compounds. This is contrary to our expectation. The softening of the rest of the Fe and As vibrations is found to be responsible for the softening of the phonon spectra below 22 meV as seen in $CaFe_2As_2$ (Figure 3). As expected the Fe-As stretching modes, around 32 meV in the Ca compound, are shifted towards higher energies due to shorter Fe-As bond length in $CaFe_2As_2$ compared to the other compounds.

For the 1111 cases, the Ca and Sr vibrations in SrFeAsF and CaFeAsF scale approximately with the mass ratio. We find that there is a significant difference in the atomic vibrations of F. The partial density of states of F atoms in both compounds extends up to full spectral range of 45 meV. The structure of these compounds has FeAs4 and F(Sr/Ca)4 tetrahedral units. The mass normalization shifts Sr vibrations down to lower energies which, in turn, affects the F vibrations. In particular a peak at 15 meV in the F partial density of states in CaFeAsF is shifted to 12 meV in SrFeAsF. The reason that mass and unit cell contraction effects are not able to explain the difference in phonon spectra may be due to the fact that the unit cell contraction in these compounds is accompanied by a substantial change [3, 37] in the free structural parameter of the As atom.

Another interesting feature, which we found by comparing the partial density of states of both $SrFe_2As_2$ and SrFeAsF, is that Fe and As vibrations are nearly the same in both the systems. This indicates that Fe-As layers are not significantly modified by additional Sr-F interaction in SrFeAsF. However the range as well as the spectrum of Sr vibrations is completely different in both compounds indicating that bonding of the Sr atoms is quite different in $SrFe_2As_2$ and SrFeAsF. This may be due to additional Sr-F interactions in SrFeAsF and consequently the interaction between the Sr and Fe-As layer is also quite different. The valence-band electrons, close to the Fermi surface, are mainly believed to be involved in the superconductivity. Electronic structure calculations for SrFeAsF show [48] that bands near the Fermi level are mainly formed by Fe 3d states. However, in the case of $SrFe_2As_2$, bonding and interaction of both Fe and As are believed [49] to be responsible for superconductivity.

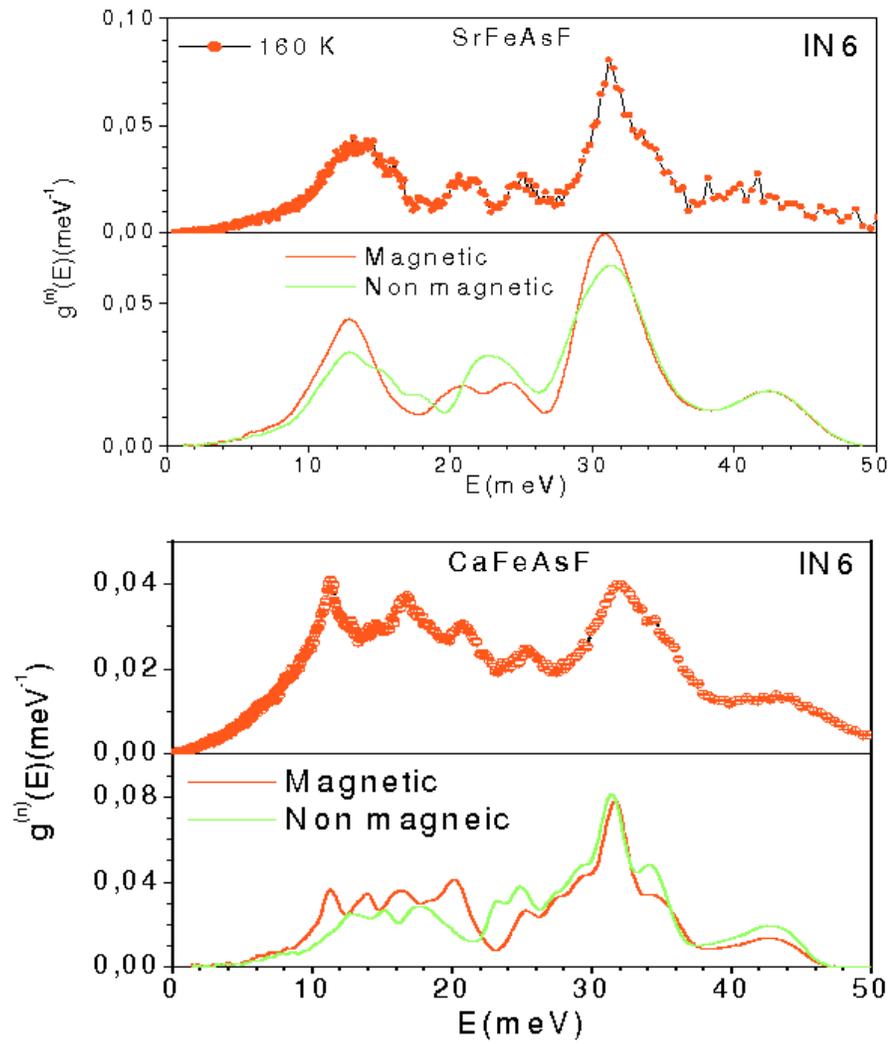

**Figure 7.** Comparison between the experimental and calculated phonon spectra for AFeAsF (A=Sr, Ca). The calculated spectra have been convoluted with a Gaussian of FWHM of 10% of the energy transfer in order to describe the effect of energy resolution in the experiment. All the phonon spectra are normalized to unity. The experimental phonon data for CaFeAsF are taken from Ref. [28].

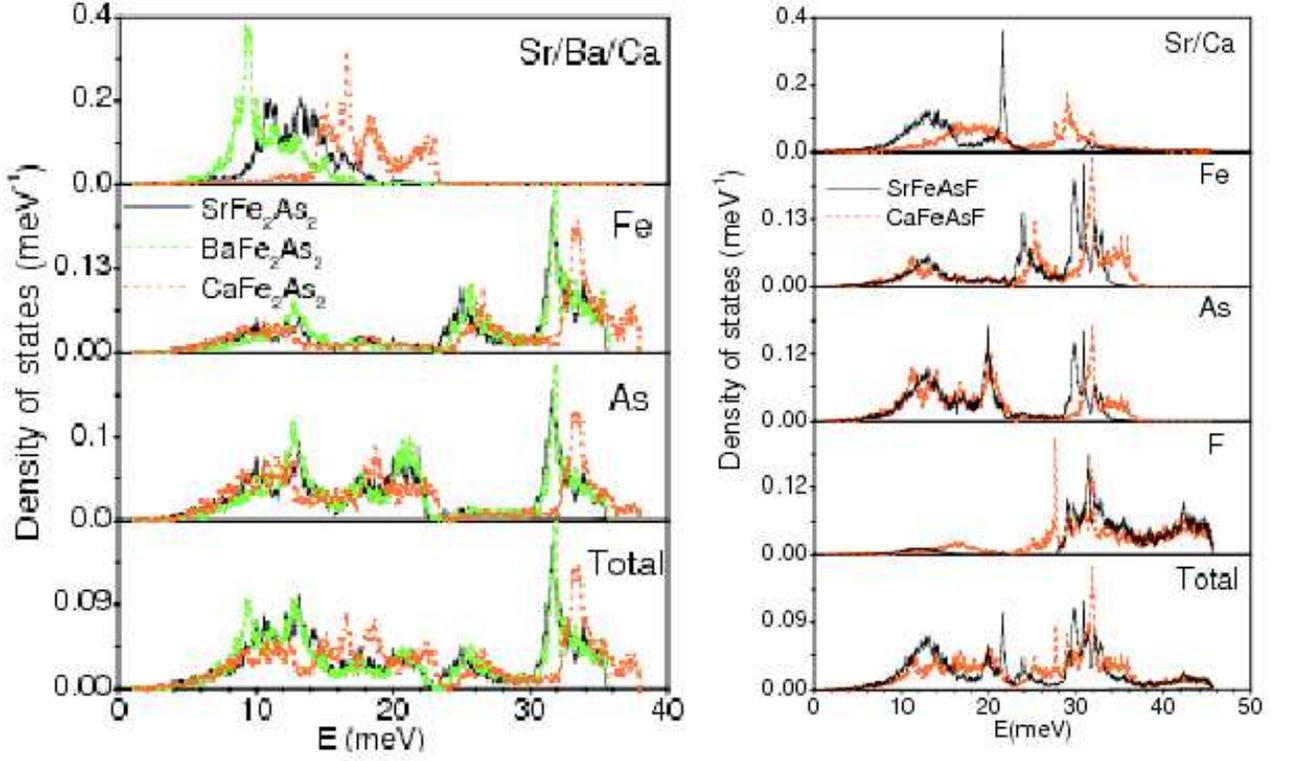

**Figure 8.** Ab initio magnetically calculated partial density of states for the various atoms in AFe$_2$As$_2$ (left panel, A=Sr, Ba, Ca) and AFeAsF (right panel, A=Sr, Ca). The spectra are normalized to unity.

## 4.4. Magnetic Ordering and Calculated Phonon Dispersion Relations in Sr systems

We have established that magnetic interactions are needed to capture the relevant features of the lattice dynamics. Indeed the agreement between calculations and measured phonon spectra is significantly improved and depends on the magnetic ordering and the related structural correlations. The differences found in the calculated phonon density of states from NM and MAG approaches can be understood via the calculated phonon dispersion relations (PDR) in the NM and MAG calculations.

The Ca and Ba cases were already reported [27, 28, 29] , here we show calculated PDR for the Sr compounds, see Figure 9. We observe that more or less flat phonon branches in SrFe$_2$As$_2$ around 24 meV and 18 meV in the NM calculations are mainly shifted to 22 meV and 16 meV in the AVE calculations (Figure 9 left panel). The calculated partial density of states (Fig. 8) shows that modes round 21 meV mainly correspond to the Fe modes while 16 meV vibrations are due to As atoms. As a result, the greater density of phonon branches in magnetic calculations for energies around 21 meV, gives rise to a peak in the spectrum (Fig. 5 upper panel) from the magnetic calculation just above 20 meV, in good agreement with the measured spectrum.

Our calculation for SrFeAsF shows that, there is a greater density of phonon branches (Figure 9 right pannel) in NM calculations for energies around 15 to 20 meV. The inclusion of magnetic structure in the calculations shifts all these branches to about 15 to 16 meV at the zone boundary. Further the flat optic modes between 19 to 22 meV are dispersive in the magnetic calculations. We observe that there is a greater density of modes around 12 meV in the magnetic calculations. The density of states measurements is mainly sensitive to zone boundary modes. This increase in density of modes at the zone boundaries results in an increase of intensity in the density of states for the peak around 12 meV.

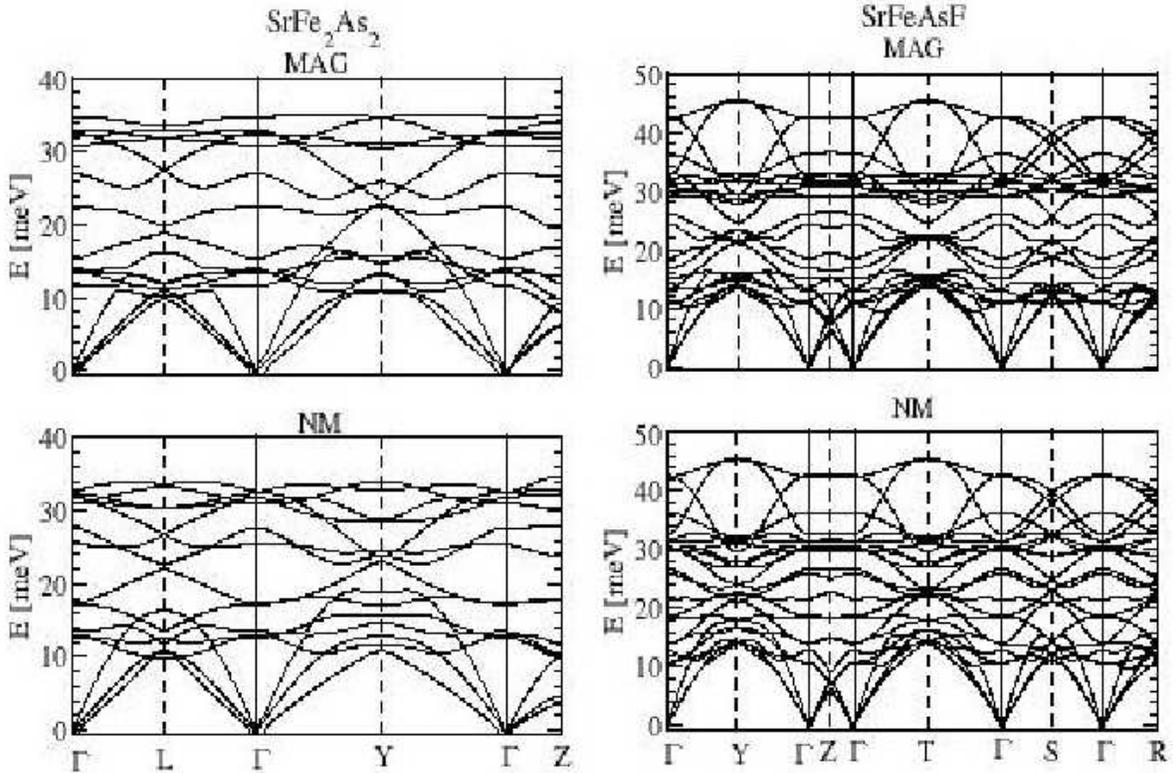

**Figure 9.** Ab initio calculated phonon dispersion relations for SrFe$_2$As$_2$ (left pannel) and SrFeAsF (right pannel). The Bradley-Cracknell notation is used for the high symmetry points along which the dispersion relations are obtained. For SrFe$_2$As$_2$: L = ( 1/2 , 0, 0), Y = ( 1/2 , 0, 1/2 ) and Z = ( 1/2 , 1/2 , 0). For SrFeAsF: Y= ( 12 , 1/2 , 0), Z=(0, 0, 1/2 ), T=( 1/2 , 1/2 , 1/2 ), S=(0, 1/2 , 0) and R=(0, 1/2 , 1/2 ).

## 4.5. Discussion: symmetry reduction in MAG phonon calculations

We have carried out additional numerical investigations of the effect of the magnetic order breaking the lattice symmetry and hence lowering the space group. We have reported this recently [28] and it has also been suggested by other authors [36]. We have done therefore the same MAG calculations as above but with two crystallographically inequivalent Fe sites to separate spin up and spin down components and mimic antiferromagnetism along the a- and c axes. We have referred previously to this procedure as the broken symmetry approach (BS) and the presently reported MAG calculations as the magnetically spin averaged approach (AVE) [28]. We do not report here the corresponding results because the AVE (MAG) and BS magnetically calculated phonon spectra are closely similar. In fact they only differ due to small geometrical changes arising from the structural optimization in a lower space group. In principle BS and AVE (MAG) approaches give identical results for the cases studied here because the pair-wise magnetic interactions are equivalent for crystallographically equivalent magnetic sites, even though the spin polarization on these sites is different. To illustrate this remark, in a fully anti-ferromagnetic structure, a spin-up polarization on a magnetic site surrounded by spin-down nearest neighbours, gives rise to the same magnetic interactions as a spin-down site surrounded by spin-up neighbours, provided the two sites are crystallographically equivalent. This argument holds for the mixed ferro (along b) and anti-ferromagnetic order (along a and c axes) in the pnictides studied here. Experimentally, the AVE (MAG) approach corresponds to a correct description of the magneto-lattice symmetry.

## 5. Summary

We have presented a detailed investigation of the effect of magnetic ordering/interactions on the phonon dynamics in the 122 - $AFe_2As_2$ (A=Ba, Sr, Ca) and 1111 - AFeAsF (A=Sr, Ca) parent compound pnictides based on ab initio lattice-dynamics simulations. In two approximations the magnetism has been neglected (NM) and then included (MAG) when calculating the inter-atomic force constants. The calculations are accompanied by new inelastic neutron scattering measurements of the temperature dependence of phonon spectra for the parent compounds $SrFe_2As_2$ and SrFeAsF as well as with our previous measurements for Ca and Ba compounds [27, 28, 29]. We show that magnetic interactions (and ordering) (MAG) have to be considered explicitly in order to reproduce correctly the measured density of states for all the presented systems. The strong sensitivity of the force constants upon inclusion of magnetism demonstrates a pronounced spin-lattice coupling which is also well reflected in some dynamical feature like the axial $A_{1g}$ mode at 21 meV of the As atoms in $BaFe_2As_2$ which is computationally reproducible only if magnetism is combined to the lattice dynamics. We demonstrate also in these systems that there is no need to reduce the crystallographic symmetry when including the effect of magnetic interactions in the force constants.

Further, the comparison of the phonon spectra in the different systems shows that as far as dynamics is concerned phonon spectra of Ba and Sr systems mimic each other. However there are substantial differences when we compare the phonon spectra of Ba and Sr compounds with the Ca system. The mass and lattice contraction effects alone cannot explain these changes. These differences may be due to the fact that the unit cell contraction is accompanied by a substantial change in the free structural parameter of the As atom. Further we have shown that in the two Sr compounds the vibrational contributions from both Fe and As are similar, whereas the vibrations from Sr are quite different, which indicates that bonding of the Sr atom is quite different in $SrFe_2As_2$ and SrFeAsF. This is due to the additional Sr-F interactions in SrFeAsF and consequently the interaction between the Sr and Fe-As layer is also quite different compared to $SrFe_2As_2$.